\documentclass[journal]{IEEEtran}%, compsoc
\usepackage{algorithm}
\usepackage{algorithmic}
\usepackage{amsmath,epsfig}
\usepackage{amssymb,float}
\usepackage{cite}
\usepackage{color}
\usepackage{bm}
\usepackage{multirow}
\IEEEoverridecommandlockouts %this for adding \thanks{}

\begin{document}

\title{ Joint Protection Scheme for Deep Neural Network Hardware Accelerators and Models
%Obfuscation Scheme for Deep Neural Network Accelerators
}

\author{\IEEEauthorblockN{Jingbo Zhou and Xinmiao Zhang,~\IEEEmembership{Senior Member,~IEEE}}\\

\thanks{The authors are with The Ohio State University, Columbus, OH 43210, USA. Emails: \{zhou.2955, zhang.8952\}@osu.edu.}}

\IEEEtitleabstractindextext{%
\begin{abstract} Deep neural networks (DNNs) are utilized in numerous image processing, object detection, and video analysis tasks and need to be implemented using hardware accelerators to achieve practical speed. Logic locking is one of the most popular methods
for preventing chip counterfeiting. Nevertheless, existing logic-locking schemes need to sacrifice the number of input patterns leading to wrong output under incorrect keys to resist the powerful satisfiability (SAT)-attack. Furthermore, DNN model inference is fault-tolerant. Hence, using a wrong key for those SAT-resistant logic-locking schemes may not affect the accuracy of DNNs. This makes the previous SAT-resistant logic-locking scheme ineffective on protecting DNN accelerators. Besides, to prevent DNN models from being illegally used, the models need to be obfuscated by the designers before they are provided to end-users. Previous obfuscation methods either require long time to retrain the model or leak information about the model. This paper proposes a joint protection scheme for DNN hardware accelerators and models. The DNN accelerator is modified using a hardware key (Hkey) and a model key (Mkey). Different from previous logic locking, the Hkey, which is used to protect the accelerator, does not affect the output when it is wrong. As a result, the SAT attack can be effectively resisted. On the other hand, a wrong Hkey leads to substantial increase in memory accesses, inference time, and energy consumption and makes the accelerator unusable. A correct Mkey can recover the DNN model that is obfuscated by the proposed method. Compared to previous model obfuscation schemes, our proposed method avoids model retraining and does not leak model information. 

\end{abstract}

}

\maketitle
\IEEEdisplaynontitleabstractindextext
\IEEEpeerreviewmaketitle

\section{Introduction}
\IEEEPARstart{D}{eep} neural networks (DNNs) are popular in numerous image processing, object detection, and video analysis tasks. Scaling up the model size can increase the learning ability and performance of a DNN model \cite{Resnet, network express}. However, large amount of data needs to be stored during DNN inference. Accessing data from memory consumes much higher energy and takes much longer time than arithmetic computations, especially when the intermediate data is big and needs to be stored in off-chip memories. Many DNN accelerators \cite{Eyeriss, Sparten, PermDNN, GOSPA, ESCALATE} realize that there is high sparsity on the temporary result of different layers in DNN inference. In order to reduce the memory size and access, these DNN accelerators discard zero values and only store non-zero values in compressed formats in memory during the model inference. 

Logic locking \cite{RoyEPIC} is one of the most popular methods to prevent chip counterfeiting. It inserts a key-controlled block into the circuit so that signals in the circuit are flipped when a wrong key is used. The correct key is handed out to the authorized user after chip fabrication. The satisfiability (SAT) attack \cite{SAT} is a powerful attack against logic locking. It iteratively excludes wrong keys corrupting the output of the locked circuit. The earlier logic-locking schemes, such as those in \cite{RoyEPIC, RajendranSecurity, AND logic, Fault analysis, Lee-logic, LUT2}, are subject to the SAT attack. The Anti-SAT \cite{Anti-SAT}, SARLock \cite{SARLock}, and G-Anti-SAT \cite{G-Anti-SAT} schemes make the number of iterations needed by the SAT attack exponential. However, in these SAT-attack-resistant schemes, the number of input patterns leading to wrong outputs under incorrect keys, which is referred to as the corruptibility of the wrong key, is small. The AppSAT attack \cite{AppSAT} excludes those high-corruptibility wrong keys and return an approximate key, which can be used to generate the correct output for most input patterns. Although the stripped functional logic locking (SFLL) \cite{SFLL} improves the trade-off between the SAT attack resistance and the corruptibility of all wrong keys, an approximate wrong key still does not affect the accuracy of DNNs much since they are naturally fault-tolerating \cite{ML}. Besides, by analyzing the functional and structural properties of SFLL, the functional analysis on logic locking (FALL) \cite{FALL} and Hamming distance (HD)-unlock method \cite{HD-unlock} can successfully compromise the SFLL. As a result, it is essential to develop an effective logic-locking method to protect DNN accelerators.

Training a DNN model requires massive labeled data, computation resource, and significant work on tuning the hyper parameters. In the white-box setting \cite{ML-doc}, the neural network architecture and the trained model parameters are publicly known (e.g. Caffe Model Zoo). In this case, the attacker can modify the trained parameters without affecting the output accuracy and claim the ownership. 
\cite{protection1} proposes a key-dependent training process to lock the original model. The target hardware platform is modified so that the original model can be recovered by providing a correct key. However, this obfuscation method requires the model to be retrained for every correct key that is sent to different users. \cite{protection2} swaps the rows and columns of the filters and/or the order of the layers of the trained model. These information is sent to the authorized end-user. Nevertheless, it can also help the attacker to recover the original model.

This paper proposes a joint logic locking and obfuscation scheme that protects both DNN hardware accelerators and models. Most state-of-the-art DNN accelerators are aware of the sparsity in the intermediate results and have data compression blocks that only record the nonzero values in order to reduce the memory requirement. By inserting a logic-locking block controlled by a hardware key (Hkey) into the data compression block, our design substantially increases the amount of data to be stored into the memory when a wrong Hkey is applied. Accessing more intermediate data stored in memory also leads to longer latency and higher energy consumption. Different from previous logic-locking schemes, using a wrong Hkey in the proposed design does not affect the circuit output. Since the SAT attack can only exclude keys corrupting the output, wrong Hkeys in the proposed design cannot be excluded, and the SAT attack can be effectively resisted. The proposed logic-locking scheme can be used by different accelerators regardless of the data compression method. The DNN model obfuscation method proposed in this paper changes the biases of a trained DNN model. The DNN accelerator is also modified to have a model key (Mkey) inserted into the bias addition block. Only when the correct Mkey determined by the DNN provider is used, the original biases are recovered. Similar to previous DNN model protection schemes, the proposed method makes the original model hard to recover even if the attacker has access to part of the training sets. The advantage of our proposed scheme is that it does not require the model to be retrained for different correct keys. More importantly, it does not leak any information that can help with the recovery of the original model.

Experiments are done on popular DNN models, such as AlexNet \cite{Alexnet}, ResNet \cite{Resnet}, and VGG \cite{VGG}, to verify the advantages of the proposed protection scheme. Simulation results using the ILSVRC-2012 dataset \cite{dataset} show that a DNN accelerator obfuscated by our proposed logic-locking scheme requires as much as 3.71, 1.67, and 2.76 times more memory accesses for AlexNet, ResNet, and VGG, respectively, with different compression methods under wrong Hkeys. In addition, without the correct Mkey, the top-1 accuracy for those three networks drops by 86\%, 76.72\%, and 82.35\%, respectively, on CIFAR-10 dataset \cite{CIFAR-10}. Even if a more powerful attacker obtains $10\%$ of original training set as thief dataset to finetune the obfuscated model, the accuracy of the finetuning result still drops by as much as $14.81\%$.

This paper is organized as follows. Section II introduces DNN hardware accelerators, logic locking, and DNN model protection. Section III proposes our new methods for jointly protecting DNN hardware accelerators and models. Experimental results and security analyses are provided in Section IV. Discussions and conclusions follow in Section V and VI, respectively.

\section{Backgrounds}
This section first introduces background knowledge of DNN accelerators. Then the threat models are presented. Previous logic locking and DNN protection schemes are also introduced in this section.

\subsection{Sparsity-aware DNN accelerators}
A DNN typically consists of different layers, such as convolutional layer, pooling layer, and fully connected layer, connected in sequence. A convolutional layer is used to extract the features from input data. A pooling layer is responsible for solving the potential overfitting problem of DNNs by reducing the spatial size of input feature maps. A fully connected layer is usually used when the DNN is designed for classification tasks. It computes the class scores based on the outputs from the previous layer. In order to increase the non-linearity of DNNs, an activation function usually follows a convolutional or fully connected layer. The rectified linear unit (ReLU) is the most popular activation function, and it is defined as
\begin{equation}
\label{relu}
ReLU(x)=\left\{
\begin{aligned}
& 0, (x < 0) \\
& x, (x \geq 0).
\end{aligned}
\right.
\end{equation}

Since modern DNNs are deep and many layers are wide, a hardware accelerator typically consists of a number of computation units that are used in a time-multiplexed way to complete the calculations layer by layer. For each convolutional and fully connected layer, the input feature map first goes through multiplier and accumulators (MACs). Then the sums of the products go through the bias adders to be added with the corresponding biases. The results of bias adders will be sent to ReLU function. The outputs of the ReLU function, called output feature maps, need to be stored before they are consumed by the computations for the next layer. 

Due to the wide layers, storing all the data in the feature maps requires large memory and many memory accesses. According to \cite{ESCALATE}, accessing an 8-bit data from DRAM costs around 200x energy compared to an MAC operation. In addition, accessing data from off-chip memory needs much longer latency despite possible pre-fetching. From \eqref{relu}, a large portion of the elements in the output feature maps are zero. Take the first convolutional layer of AlexNet as an example. When 128 images from the ILSVRC-2012 dataset are used as the inputs, 60.95\% of values in the output feature maps are zero. To reduce the amount of data to store, different compression methods, such as compressed sparse column (CSC) format \cite{GOSPA}, run length compressing (RLC) \cite{Eyeriss}, and bit map format \cite{Sparten, ESCALATE}, have been utilized in sparsity-aware DNN accelerators. These compression methods discard zeros. Non-zero values and their locations are recorded in different formats.

\subsection{Threat models}

\begin{figure*}[h]
    \centering
    \includegraphics[width=6.0in]{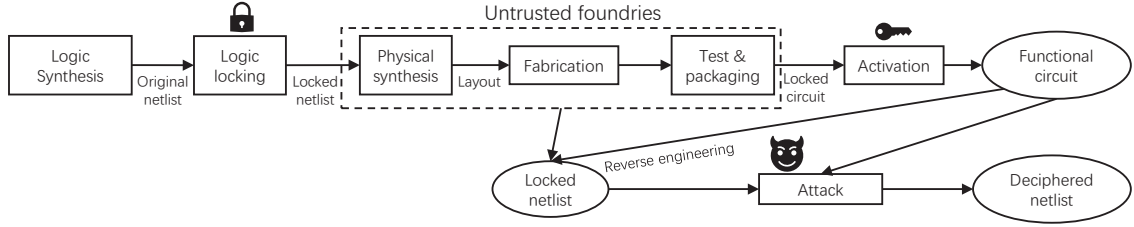}
    \caption{Models of hardware threats}
    \label{attack_model_hardware}
\end{figure*}
\subsubsection{Hardware threat model}

The hardware threat model is shown in Fig. \ref{attack_model_hardware}. Many of the circuit designs are sent to a third-party facility for fabrication. If the attacker gets the netlist, extra chips may be made and sold for profit. Such chip overproduction causes significant economic loss to integrated circuit design companies. In an integrated circuit supply chain, an attacker can get the netlist of the circuit design either from the untrusted foundry or by reverse engineering a chip bought from the open market. Following the same assumption made by previous papers, the attacker has access to: 1) the netlist of the chip; 2) a functional chip that can be purchased from the open market. However, the attacker does not have access to: a) the intermediate signals of the circuit since they cannot be probed; b) the correct key stored in temper-proof memory. 

\begin{figure*}[h]
    \centering
    \includegraphics[width=6.0in]{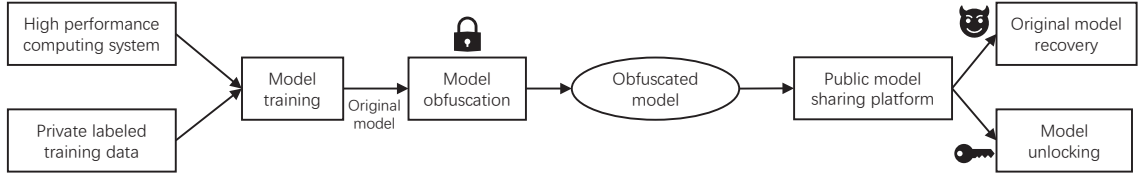}
    \caption{Models of DNN threats}
    \label{attack_model_DNN}
\end{figure*}
\subsubsection{DNN threat model}
Training a DNN model requires significant work including designing the model structure, labeling the training and validation dataset, and tuning the hyperparameters. In addition, a large amount of computing resource is needed to complete the training in practical time. To prevent a trained model to be illegally used, it needs to be obfuscated to make the output accuracy drop significantly under unauthorization usage. The goal of the attacker is to recover the original model and distribute the model on his/her own systems.

In this paper, the attacker is assumed to have access to the structure and parameters of the obfuscated model as shown in Fig. \ref{attack_model_DNN}, just like the white-box setting mentioned in \cite{ML-doc}. In addition, a small portion, $\it i.e.$ 10\%, of the private labeled training data is accessible according to \cite{protection1}. Utilizing the small portion of training data as well as the structure and obfuscated parameters of the DNN, the attacker can finetune the DNN to increase the accuracy. 

\subsection{Related work}

\subsubsection{Existing logic-locking schemes and attacks}
\begin{figure}[t]
    \centering
    \includegraphics[width=2.5in]{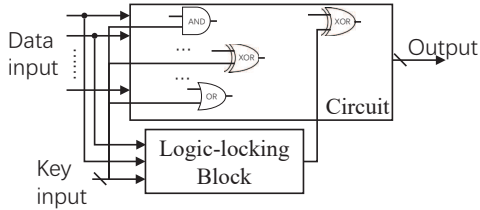}
    \caption{Circuits protected by logic locking}
    \label{logic locking}
\end{figure}

Logic locking inserts key-controlled blocks or introduces individual key bits into the circuit by using XOR/XNOR/AND/OR gates as shown in Fig. \ref{logic locking}. When a wrong key is used, intermediate signals of the circuit are flipped and the circuits output is corrupted under certain input patterns. The SAT attack \cite{SAT} is one of the most powerful methods that can compromise logic locking. It utilizes the conjunctive norm formula of the locked circuit netlist. By carrying out queries to the functional chip, wrong keys that make the output of the locked circuit different from the original output are excluded by the SAT attack. To resist the SAT attack, many logic-locking schemes have been developed, such as the SARLock \cite{SARLock} and Anti-SAT \cite{Anti-SAT}. Although they make the number of iterations needed by the SAT attack exponential to the number of key input bits, the corruptibility of each wrong key is only one. This means that using any wrong key will not corrupt the output most of the time. The low corruptibility is utilized by the AppSAT attack to return an approximate key. 

The G-Anti-SAT \cite{G-Anti-SAT} design maintains the exponential iterations in the SAT attack. On the other hand, it substantially increases the corruptibility of a large portion of the wrong keys. A variation of the G-Anti-SAT was also developed in \cite{ISCAS} to better resist removal attacks. The SFLL proposed in \cite{SFLL} increases the corruptibility of all wrong keys at the cost of reduced number of iterations needed by the SAT attack. However, none of existing logic-locking schemes can effectively protect hardware chips implementing DNN inference since it is fault-tolerant \cite{DNN reliability}. Even if intermediate signals in a DNN accelerator are locked by SFLL, the output accuracy is not affected when a wrong key is utilized as it was found in \cite{ML}. 

\subsubsection{Previous methods protecting DNN model}
A trained model needs to be obfuscated before being provided to end-users. Authorized users can inference the model without accuracy degradation. \cite{protection1} proposes a key-dependent back propagation algorithm to train DNN models. Random neurons from a DNN model are chosen. They are modified according to the selected key vector and the data scheduling scheme of the target hardware accelerator. Retraining is carried out on the modified model and the weights are changed. The hardware accelerator implementing the obfuscated model is also modified so that only the correct key can unlock the model. 
However, for every different correct key, this obfuscation method requires the model to be retrained. Retraining a model is computation intensive and requires long time. Hence, the scalability of this method is limited.

\cite{protection2} proposes a method that does not need retraining. It obfuscates the model by swapping the rows, columns, and layers of the weight matrices. The correct key indicating the information about the swapping is handed out to the authorized end user by the model designer. Based on these information, the end user can get the original model back and make the correct inference. Without the right key, the weight matrices are swapped in the inference and the accuracy of the modified model drops significantly. Nevertheless, if the key itself gets compromised, the original model will be recovered and can be illegally used.

\section{Protection Schemes for sparsity-aware DNN Accelerators and Models}
This section proposes an effective joint protection scheme for DNN hardware accelerators and models. The proposed method inserts an Hkey-controlled block after the ReLU function so that zero values will not be recognized under incorrect Hkeys. As a result, all values go through the compression process and are stored into memory. In this case, the latency and power consumption of the hardware accelerator are increased by multiple times. In addition, the ReLU function is modified to further prevent the Hkey-controlled block from being bypassed. DNN models are protected by introducing an Mkey in our design. The biases of the DNN model are obfuscated and the bias adders in the hardware accelerator implementing the obfuscated model is modified to have an Mkey input. Only the correct Mkey can recover the correct biases and make DNN inferenced without accuracy degradation.

\subsection{Hkey-controlled match detector and modified ReLU}
Let the output of the ReLU function be $X$. Existing sparsity-aware DNN hardware accelerators discard $X$ if it is zero and only store the non-zero values in compressed format in the memory. In the proposed design, the zero detector in existing designs is replaced by a Hkey-controlled match detector. Under the correct Hkey, it outputs `1' if $X$ is zero and '0' otherwise. The following data compression unit compresses $X$ when the match detector outputs '0'. Wrong Hkeys make the detector always output '0', and all values including zeros will go through the compression block and get stored into memory.

The functionality of the ReLU is public and it is well-known that sparsity-aware DNN hardware accelerators discard the zeros. A potential threat is that the output wires of the ReLU function may be located by the attacker. Then, a zero detector can be used to bypass the match detector and nullify the inserted Hkey. To prevent such threats, the ReLU function is modified in our design. Assume that the input to the ReLU is an $h$-bit value $A = a_{h-1}\dots a_1a_{0}$ and $a_{h-1}$ is the sign bit. Then the output of the original ReLU defined in \eqref{relu} can be described in logic level as 
\begin{equation*}
    x_{h-1} = 0, x_j = \overline{a_{h-1}} \& a_j (0\leq j<h-1),
\end{equation*}
where the overhead bar denotes logic NOT. In our modified ReLU function, a pre-determined constant $h$-bit secret vector $T=t_{h-1}\dots t_1t_{0}$ is XORed with $X$ to generate $X'$ as
\begin{equation}\label{at}
    x'_{h-1} = t_{h-1}, x'_{j} = x_j\oplus t_j(0\leq j<h-1).
\end{equation}
When $t_j=$'1' and '0', $x'_j=\overline{x_j}$ and $x_j$, respectively. Hence, depending on the bit pattern of $T$, there is either an extra NOT gate or no extra NOT gate on the bits of $X$. The NOT gates are combined with other logic by synthesis tools and are not easily discernible from the netlist. Therefore, the secret $T$ vector can not be recovered from the netlist. Accordingly, even if the output wires of the modified ReLU function can be located, the attacker can not replace the match detector with a zero detector to correctly recognize the zero values.

Using a modified ReLU function, whose output is $X'$, the functionality of the proposed match detector controlled by Hkey, $HK$, is
\begin{equation}\label{logic}
\begin{aligned}
g(X', HK) & = f_k(HK) \& f_x(X'). \\
\end{aligned}
\end{equation}
$f_x(X') = \tilde x_{h-1} \& \tilde x_{h-2} \& \dots \tilde x_0$ and $\tilde x_j = {x'_j}$ XNOR $t_j$. Hence, $f_x(X')$ detects whether $X$ is zero. Assume that $HK$ has $c$ bits and the correct $HK$ determined by the designer is $HK^*$. The function $f_k(HK)$ is $\overline{hk_{c-1} \oplus hk^*_{c-1}} \& \dots \& \overline{hk_{0} \oplus hk^*_{0}}$. Basically, $f_k(HK)$ is '1' only when the input $HK$ equals the correct Hkey, $HK^*$. When $HK=HK^*$, $g(X', HK)= f_x(X')$ and it is '1' when the output of the original ReLU function is zero. Otherwise, $g(X', HK)$ is always '0', and all of the values, including zero values, will be sent to the compression block. $HK^*$ is secret and its bits complement some of the bits in the $HK$ input. The NOT gates at those bits of $HK$ are merged with the AND functions of $f_k(HK)$ and $f_x(X')$. As a result, the $f_k(HK)$ function cannot be separated from the netlist, and the Hkey cannot be nullified by replacing $f_k(HK)$ with constant signal '1'.

\subsection{Hardware-assisted DNN model protection}
%\begin{figure}[t]
  %  \centering
    %\includegraphics[width=2.5in]{DNN_IP_protection.eps}
    %\caption{Bias addition block and that block with Mkey inserted}
    %\label{Mkey insertion}
%\end{figure}

In the white-box setting \cite{ML-doc}, a trained neural network architecture and its parameters are publicly known. To prevent attackers from illegally using the trained model in their own systems, obfuscation should be made on the model parameters by the designer before publishing the model on public platforms. In a DNN model, weights and biases are two sets of parameters affecting the accuracy. Corrupting either of them can make the model accuracy drop significantly. Our design chooses to obfuscate the biases since it also helps to better prevent bypassing the match detector for logic locking proposed in the previous subsection. It is easy to identify the registers used to store the MAC results and memories for recording the output feature maps from the netlist. The match detector is a combinational logic part between the registers and memories and hence is more vulnerable to be isolated and being compromised by removal attacks. By obfuscating the biases, even if the registers and memories are identified, the correct function of the logic part between them can not be restored without knowing the modifications made on the bias adders. As a result, the Hkey-controlled match detector cannot be bypassed.

From our simulations, it was discovered that the output accuracy of the DNN models is significantly degraded if at least the two most significant bits (MSBs) of every bias are modified. Utilizing this observation, our proposed scheme XORs a random-selected vector to the two MSBs of biases and provide these modified biases to the model sharing platform. In the DNN accelerator of the authorized end user, the correct Mkey is used to XOR obfuscated biases back for the bias adders.
There are many biases in a DNN model. However, the Mkey should have limited length. In a DNN hardware accelerator, although multiple MAC blocks and bias adders are utilized to increase the throughput, their numbers are much smaller than those of the weights and biases in a DNN. Hence, the adders need to be shared in a time-multiplexed way to add different biases over a number of clock cycles according to the data scheduling scheme. Hence, the model provider should use the same vector to modify the biases sharing the same adders over different clock cycles according to the data scheduling scheme of the DNN accelerator. Accordingly, the correct Mkey can be used for a bias adder for different clock cycles to restore the biases sharing that bias adder.

\subsection{Hardware implementation architecture}

\begin{figure*}[t]
    \centering
    \includegraphics[width=6.5in]{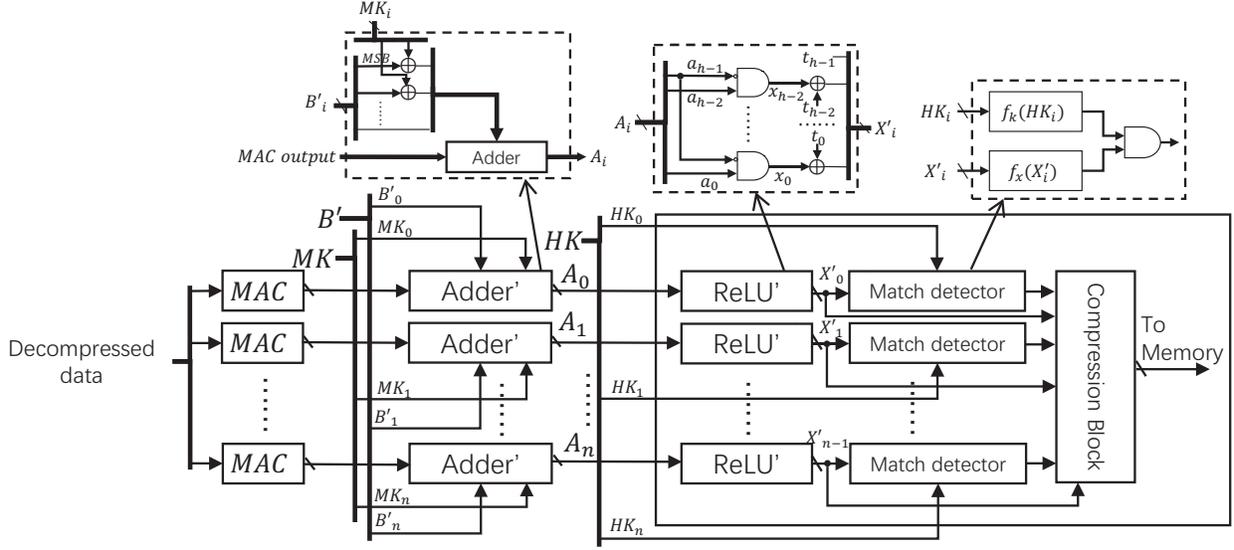}
    \caption{Overall architecture of DNN hardware accelerator with the proposed joint protection scheme.}
    \label{Hkey insertion}
\end{figure*}

The architecture of the DNN accelerator incorporating the proposed joint protection scheme is shown in Fig. \ref{Hkey insertion}. Multiple MACs are utilized to achieve high throughput, and the same number of bias adders, ReLU function blocks, and match detectors follow. The implementation of the modified bias adders, ReLU function blocks, and match detectors are summarized in this subsection.  

As mentioned in the previous subsection, at least the first two MSBs of each bias is XORed with the randomly-chosen secret vector. In the bias adders of the hardware accelerators, biases are XORed back by using correct $MK$. In Fig. \ref{Hkey insertion}, the segment of $MK$ sent to the $i$-th bias adder is denoted by $MK_i$. In the case that there is a very large number of bias adders, a portion of them can be modified to insert $MK$ in order to limit the length of $MK$, which means part of biases in a DNN model are not obfuscated, although the degradation on the model accuracy might be less significant. To further protect the DNN model, some random XORs can be replaced by XNORs in the modified biases adders and the correct $MK$ is changed accordingly. The XOR and XNOR gates are difficult to discern from the netlist. In this case, even if the correct $MK$ is recovered, the original biases are still unknown.

In our design, the ReLU function is modified to have the output XORed with a random constant vector $T$. The $T$ vector needs to be added back when the output feature map is read from the memory for processing the next layer of the DNN. Using a different $T$ vector for each ReLU block makes it difficult to recover the original feature map. Hence, the same $T$ vector is used in every modified ReLU unit in our design. 

In our match detector, the output of the $f_x$ function is the same as that of a normal zero detector attached to an original ReLU function. If the $HK$ input matches the correct key, $HK^*$, then the detector result is passed intact to the compression block and the zero values are discarded. Otherwise, every data goes through the compression block and gets stored into the memory. In Fig. \ref{Hkey insertion}, the segment of $HK$ sent to the $i$-th match detector is denoted by $HK_i$. Similarly, to limit the length of $HK$, some match detectors can be replaced by normal zero detectors without any $HK$ input, although the increase in the memory requirement will be less significant. It should be noted that a wrong $HK$ increases the memory needed to store intermediate data without affecting the output of the DNN. More detailed discussions on the achievable security level of our protection scheme are provided in the next section.

\section{Experiment Results and Security Analyses}
This section carries out experiments and analyzes the effectiveness of the proposed joint logic locking and DNN model protection scheme.

\subsection{Security analyses of the proposed logic locking}
The main goal of hardware attackers is to utilize counterfeit circuits without authorization from the circuit designer. They try to derive the correct Hkey or remove the part of the circuit to nullify the Hkey. The SAT and removal attacks are the two potential major threats that may compromise the proposed logic locking. The resistance to the SAT attacks and removal attacks is analyzed in this subsection.

\subsubsection{Resistance to SAT attacks}
From Section II.C, by converting the netlist to conjunctive normal formula and querying a functioning chip, the SAT attack excludes the wrong keys that corrupt the primary output of the target circuit. After all wrong keys are excluded, any of the remaining keys can be used as a correct key. In our design, if a wrong Hkey is used, the output of the DNN does not change although the amount of intermediate data to store increases by several times.  As a result, none of the wrong Hkeys can be excluded by the SAT attack. Similarly, our proposed scheme is resilient to other attacks based on the SAT attack, such as the AppSAT attack.

\begin{figure*}[t]
    \centering
    \includegraphics[width=6.5in]{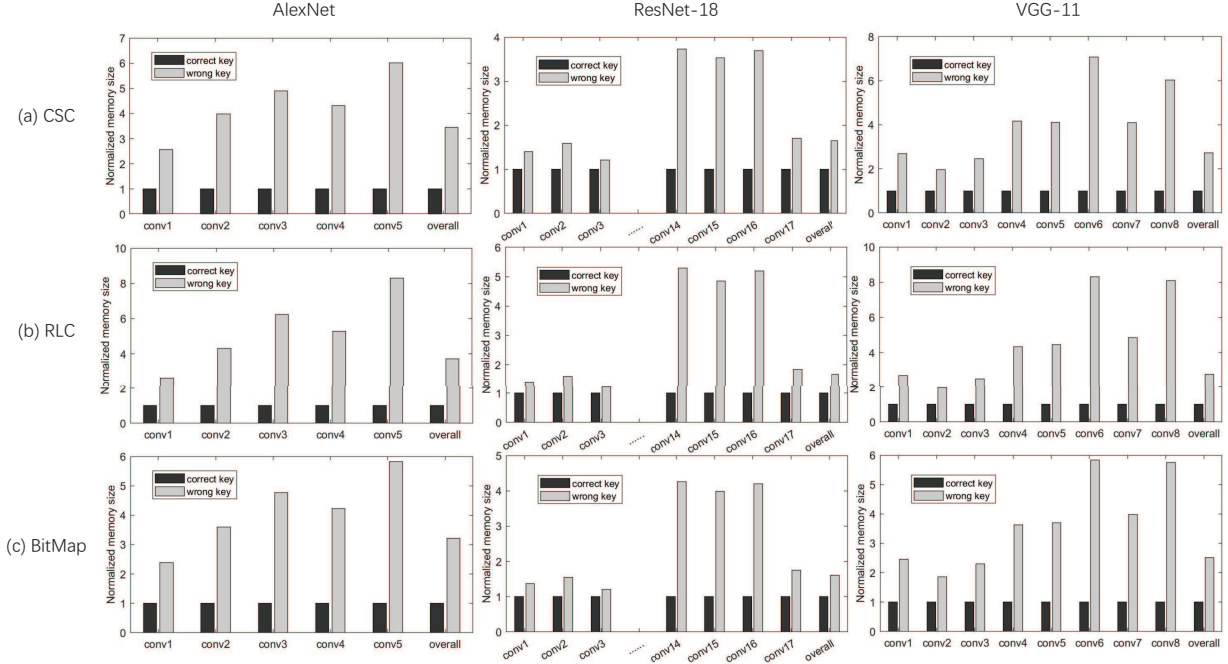}
    \caption{Memory size increase ratio caused by wrong Hkey for AlexNet, ResNet-18, and VGG-11 when the compression method is (a) CSC; (b) RLC; (c) BitMap.}
    \label{comp}
\end{figure*}

\subsubsection{Performance degradation under wrong Hkeys}
As explained in the previous paragraph, SAT attacks are not able to exclude any wrong Hkeys in our proposed scheme. If a random wrong Hkey is utilized, the amount of intermediate data to store is substantially increased as shown in Fig. \ref{comp}. In our experiments, a wrong Hkey segment is used for each match detector. To collect the data, 4 images in the ILSVRC-2012 dataset \cite{dataset} are fed to three popular DNN models: AlexNet, ResNet-18, and VGG-11. Output feature maps of each layer need to be compressed and stored into memory. Three compression formats, CSC, RLC, and BitMap, are applied to compress the intermediate data. When the Hkey segment sent to every match detector is wrong, all zero values are mistakenly considered as non-zero and sent to the compression block, which converts the data into different formats according to the compression scheme without discarding any data. The sizes of the memories needed to store the compressed data with and without zero values discarded are recorded and their ratio is computed for each layer as shown in Fig. \ref{comp}.  The overall ratio is calculated by dividing the sums of the memories required for each layer. The memory requirements for some layers are increased by more than 8 times and the overall memory requirement is increased by 1.5 to 3.7 times depending on the network and compression schemes.

The sparsity of the output feature maps from different layers of the DNN model varies. In general, if the original data has higher sparsity, then the ratio of memory increase as a result of using a wrong Hkey is larger since more zeros that should have been discarded are stored into the memory. Take the conv1 and conv2 layers in AlexNet as an example. The percentage of zeros in their output feature maps are $59\%$ and $81\%$, respectively. Hence, for different compression methods, the memory size increase ratio caused by wrong Hkey on conv2 layer is always larger than that on conv1 layer. As for the overall memory requirement increase ratio caused by wrong Hkey, ResNet-18 is smaller than the other two models. Although the memory sizes are increased more than 3 times under wrong Hkey in certain layers of ResNet-18, such as conv14, conv15, and conv16, those layers have smaller output feature maps compared to the other layers. 

Besides non-zero values, the information about their locations generated by compression also needs be recorded. Storing dense data compressed by the CSC or RLC methods requires larger memory comparing to recording the results compressed by the BitMap scheme \cite{Sparten}. When all zero values are mistakenly considered as non-zero under wrong Hkey, the original sparse intermediate results become dense. As a result, for the same DNN model, the Bitmap compression method causes the smallest memory increase ratio when wrong Hkey is utilized. Take AlexNet as an example. From Fig. \ref{comp}, the overall memory increase ratio is 3.2, 3.45 and 3.71 times, when the BitMap, CSC and RLC compression, respectively, are utilized.

Fig. \ref{comp} shows the results when the Hkey segment fed to all match detectors are wrong. However, a randomly selected $HK_i$ may be accidentally equal to $HK_i^*$. Since $f_k(HK_i)$ is a $c$-bit equality tester as discussed in the previous section, only one of the $2^c$ patterns of possible $HK_i$ is the correct key segment. Hence, the probability that a random Hkey segment for a match detector is wrong is $({2^c-1})/{2^c}$. For given length of $HK$ and number of match detectors, $n$, one choice is to evenly divide $HK$ into $n$ segments and use one segment in each of the $n$ match detectors. However, if $c$ is too small, the probability that a randomly chosen $HK_i$ equals $HK_i^*$ is higher and accordingly the memory size increase ratio is reduced. An alternative option is not to lock every match detector and evenly distribute the Hkey bits over a portion of the match detectors. In this case, the unlocked match detectors will not contribute to any increase in memory size. Simulations need to be carried out for these two options to find the better choice depending on the Hkey length and number of match detectors.

\subsubsection{Removal attack resistance}
Removal attack \cite{removal} is another hardware threat to logically locked DNN accelerators. By analyzing the netlist of the circuit, it tries to remove the logic-locking block and replace its output by a constant signal that makes the circuit function correctly. AND/NAND trees with key and data inputs as fan-ins are usually utilized to achieve SAT-attack resistance. On the other hand, large AND/NAND trees can be identified by calculating signal skews \cite{removal}. Once an AND/NAND tree is located, the removal attack replaces its output by a constant '0' or '1'. AND/NAND trees generate the outputs of the match detectors locked by Hkey in our proposed design. If the outputs are replaced by '0's, all values are passed to the compression block and stored into memory, which makes the memory requirements significantly increase as shown in Fig. \ref{comp}. If the outputs are replaced by '1's, no value is stored into memory, and the DNN accelerator output will be totally wrong and cannot be used. Therefore, our proposed design is not subject to removal attacks.

A more powerful attacker may have detailed knowledge about the architecture of the DNN accelerator, and he/she may identify the logic parts in the netlist performing certain functions. Registers and memories can be easily located in the netlist, and their inputs and outputs are identifiable. The output of the ReLU function is connected to registers in the compression block in some designs. Even if it is located by the attacker and the match detector is replaced by a zero detector, the zero values are not correctly identified since they have been XORed with the secret vector $T$ as discussed in Section III.A. The MAC units consist of registers too. Another potential attack is to replace all the logic between the MAC and compression blocks shown in Fig. \ref{Hkey insertion} by bias adders, ReLU units, and zero detectors. However, both XOR and XNOR gates can be used to connect the Mkey and obfuscated bias inputs as discussed in Section III.C. The modifications made on bias adders are secret to the hardware attacker. As a result, this attack will not be successful. 

\subsection{Security analyses of the proposed model protection}
DNN model attackers try to recover the original DNN model from the obfuscated model. Then the attackers can claim the ownership of the original model and distribute it to unauthorized users. In this section, the DNN model accuracy degradation achieved by the proposed bias obfuscation is first presented. Finetune \cite{protection1} attack may be used to recover the original DNN model under the threat assumptions introduced in Section II. B. The resistance to the finetune attack of our design is also demonstrated through simulation results in this section. 

\subsubsection{Accuracy drop on obfuscated model}
Three DNN models, AlexNet, ResNet-18, and VGG-11, with the CIFAR-10 dataset \cite{CIFAR-10} are simulated to collect the accuracy results. It is assumed that each weight, bias, and intermediate result is represented by a 16-bit fixed-point number. All the biases are divided into 128 groups in our simulations, and the two MSBs of every bias in the same group are XORed with the same two-bit random vector. Hence, the overall length of the Mkey should be 256 bits. The top-1 accuracy of the original and obfuscated DNN models are shown in Table \ref{accuracy drop}. It can be observed that obfuscating biases can effectively reduce the accuracy of DNN models. 

\begin{table*}[t]
    \centering
    \caption{The top-1 accuracy (\%) of the original DNN model, obfuscated model, and obfuscated DNN model with finetune attack applied using training sets of different sizes}
    \begin{tabular}{c||c|c|c|c|c|c|c|c|c|c|c}
         \hline
         \multirow{3}{1.4cm}{DNN models} & \multirow{3}{0.9cm}{Original model accuracy} & \multicolumn{2}{|c|}{\multirow{2}*{Obfuscated model}} & \multicolumn{8}{|c}{Obfuscated model with finetune attack applied using $\alpha$ of original training dataset} \\
         \cline{5-12}
         & & \multicolumn{2}{|c|}{~}& \multicolumn{2}{|c|}{$\alpha = 1\%$} & \multicolumn{2}{|c|}{$\alpha = 3\%$} & \multicolumn{2}{|c|}{$\alpha = 5\%$} & \multicolumn{2}{|c}{$\alpha = 10\%$}\\
         \cline{3-12}
         & & Accuracy & \% drop & Accuracy & \% drop & Accuracy & \% drop & Accuracy & \% drop & Accuracy & \% drop \\
         \hline
         AlexNet \cite{Alexnet} & 86.72 & 10.00 & 76.72 & 54.73 & 31.99 & 59.44 & 27.28 & 63.87 & 22.85 & 71.91 & 14.81\\
         \hline
         ResNet-18 \cite{Resnet} & 96.16 & 10.68 & 85.48 & 59.77 & 36.39 & 72.2 & 23.96 & 78.4 & 17.76 & 82.48 & 13.68\\
         \hline
         VGG-11 \cite{VGG} & 92.35 & 16.94 & 75.41 & 60.33 & 32.02 & 71.18 & 21.17 & 75.61 & 16.74 & 80.2 & 12.15\\
         \hline
    \end{tabular} 
    
    \label{accuracy drop}
\end{table*}

As discussed in Section II.A, DNN model attackers may have access to a small portion of the training set as thief dataset \cite{protection1}. By using the thief dataset, attackers can finetune the obfuscated model to increase the accuracy. To show the resilience of the proposed protection scheme to the finetune attack, simulations with $\alpha = 1\%$, 3\%, 5\%, and $10\%$ of the training set as the thief dataset are carried out. The thief dataset is limited to 10\%, since the labeled training data is typically kept secret by the model provider. The simulation results are shown in Table \ref{accuracy drop}. The larger the thieft dataset, the higher accuracy that the finetune attack can achieve. Nevertheless, even if $10\%$ of the original training set is available to the attacker, the accuracies of the finetuned obfuscated AlexNet, ResNet, and VGGNet are only 14.81\%, 13.68\%, and 12.15\%, respectively, which are much lower than those of the original models. 

Finetune attack has been performed on ResNet-18 obfuscated by the method proposed in \cite{protection1}. In that paper, the Fashion-MNIST \cite{Fashion} is used as the training set. When $10\%$ of the original training set is available to the attacker, the accuracy only drops by around $5\%$ compared to that of the original model. Simulation using the Fashion-MNIST dataset is also done on ResNet-18 obfuscated by the proposed design. From our simulation results, when the thief dataset is $10\%$, the accuracy of the finetuned model drops by $9.17\%$ compared to the original accuracy. This result clearly shows that the proposed obfuscation scheme is more effective than the obfuscation method in \cite{protection1} in terms of preventing finetune attack.

\subsubsection{Allocation of Mkey}
The size of Mkey, $L_M$, is decided by the model provider according to the desired security level. As mentioned previously, at least the two MSBs of the biases are obfuscated to generate significant accuracy loss in the model. Assume that there are $n$ bias adders. If $L_M>2n$, then the $L_M/n$ MSBs of each bias can be obfuscated by XORing with the bits in Mkey. When $L_M<2n$, $L_M/2$ instead of $n$ groups of biases can be chosen to be obfuscated and each bias in those groups are XORed with the bits in MKey in the 2 MSBs. As long as a significant portion, if not all, of the biases are obfuscated, the accuracy of the model may still drop significantly. Take a DNN hardware accelerator with 100 bias adders implementing AlexNet as an example. When the Mkey length is 128-bit, 64 bias adders can be chosen to be obfuscated by the $2\times 64=128$-bit Mkey. Of course, the corresponding biases in the model are modified accordingly. Without the right Mkey, the output accuracy is dropped to $10\%$, which is $76.72\%$ lower than the accuracy of the original model. This accuracy drop is similar compared to using a 256-bit Mkey to obfuscate the 2 MSBs of each bias.  

\subsubsection{Information leakage analyses}
\begin{figure}[t]
    \centering
    \includegraphics[width=3.3in]{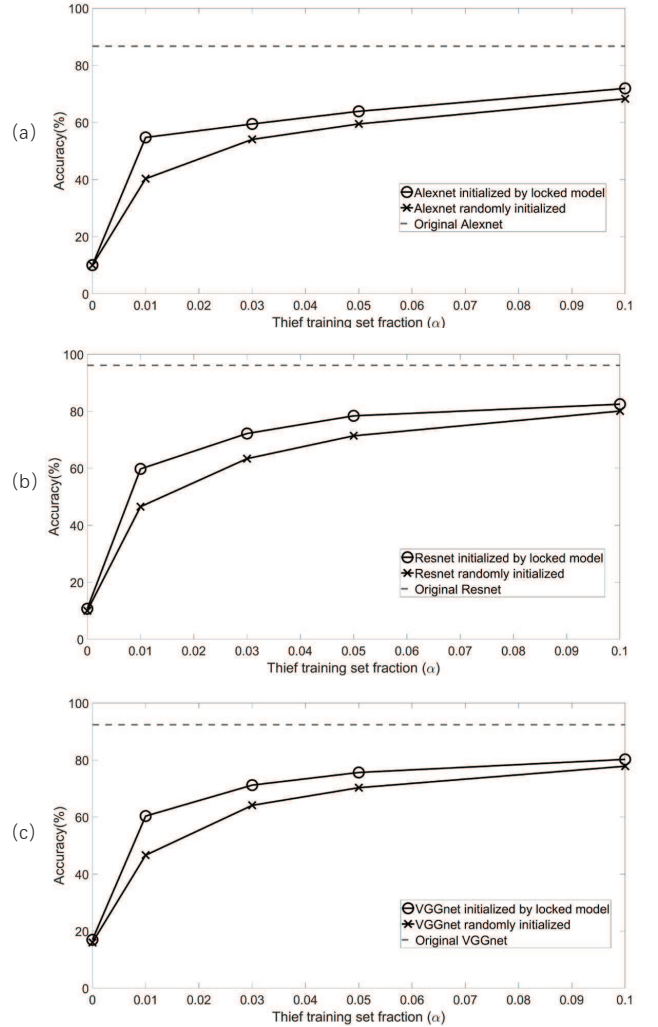}
    \caption{Finetune attack results on obfuscated models and models initialized by random parameters on (a) AlexNet; (b) Resnet; (c) VGGnet.}
    \label{Finetune}
\end{figure}

To show that the proposed DNN protection method does not leak information about the original model, the finetune attack is also applied to a model initialized by random parameters. The accuracy of the randomly initialized model after the finetune training is shown in Fig. \ref{Finetune}. The accuracies of the original model and the proposed obfuscated model after the finetune attack listed in Table \ref{accuracy drop} are also re-plotted in this figure. It can be observed that under the same size of the thief training set, finetuning the randomly initialized model and DNN model initialized by the obfuscated parameters leads to similar accuracy. More importantly, for both cases, the accuracies are much lower than that of the original model. This indicates that initializing a model with the parameters obfuscated by the proposed methods does not leak more information about the original DNN model compared to initializing the model randomly.

\section{Discussions}
This paper proposes a joint protection scheme for DNN hardware accelerator and model. The goals of the hardware and DNN model attackers are different. For the proposed protection scheme, the hardware attacker tries to derive the Hkey, which can help to unlock counterfeit chips. However, a correct Mkey does not help to recover the original DNN model, which is the goal of the model attacker. In our proposed DNN model protection scheme, the biases that are added in the same adder of the hardware accelerator are obfuscated by the same vector. Without perfect knowledge about the scheduling scheme and the netlist of the locked circuit, the model attacker does not know which bits of the Mkey are used to obfuscate each bias and how the original biases are obfuscated. Accordingly, the original model cannot be recovered. This is different from the protection scheme developed in \cite{protection2}, which hands out trusted end-users correct keys containing the indices about the swapped matrices. For that protection scheme, the attacker is able to recover the original model by using the correct key. Hence, brute-force attacks can be used to derive the correct keys and accordingly recover the original model \cite{protection2}. On the contrary, even if the correct Mkey is discovered, the model obfuscated by our proposed method still cannot be recovered. 

Since each $HK_i$ is short, one potential attack is to try all possible patterns of $HK_i$. If a certain pattern leads to noticeably smaller memory requirement, then it is the correct key segment. However, running simulations for finding the memory requirement takes very long time even if $HK_i$ is short. Additionally, it is difficult to separate the bits in one $HK_i$ from those of another. One reason is that the synthesis tool merges logic gates of different match detectors since their outputs are mixed in the calculations for compression. Besides, depending on the networks and layers, there may be other blocks, such as max pooling, in a DNN accelerator in addition to those shown in Fig. \ref{Hkey insertion}.  As a result, it is very difficult to tell the boundaries of different match detectors and separate different $HK_i$.

The hardware attacker can get the obfuscated model and the correct Mkey from the model provider. However, the correct Mkey does not help the attacker to recover the Hkey. This is because that the wrong Hkeys still cannot be excluded by SAT attacks even if the correct Mkey is available.

% As for the removal attack, the vectors used to obfuscate the original biases are secret to the attacker. Hence, even with correct Mkey, the gates used to connect the obfuscated bias and the corresponding Mkey inputs are also unknown to the attacker. Then the attacker still cannot reconstruct the parts between the MAC and the compression blocks as shown in Fig. \ref{Hkey insertion}, and the removal attack cannot success. It can be observed that the security of hardware and DNN model protection is not affected by each other.

\section{Conclusions}
In this paper, a novel scheme is proposed to jointly protect DNN hardware accelerators and models. Using the proposed scheme, the memory requirement is increased by several times when the wrong hardware key is used. This leads to substantial increases in memory size, power consumption, and latency, which make the hardware accelerator unusable. Our proposed scheme can also effectively resist SAT and removal attacks, which are among the most powerful attacks for logic locking schemes. Our proposed design protects the DNN model as well. Unlike previous approaches, our scheme is scalable and more secure.   Future research will focus on developing obfuscation schemes for DNN models and its accelerators resisting more powerful attacks.

\begin{IEEEbiography}{Jingbo Zhou}
received the B.S. degree in telecommunication engineering from Beijing University of Post and Telecommunication, Beijing, China. He is currently pursuing the Ph.D. degree in the Electrical and Computer Engineering Department, The Ohio State University, OH, USA.

His current research interest is hardware security.
\end{IEEEbiography}

\begin{IEEEbiography}{Xinmiao Zhang}
received her Ph.D. degree in Electrical Engineering from the University of Minnesota. She joined The Ohio State University as an Associate Professor in 2017. Prior to that, she was a Timothy E. and Allison L. Schroeder Assistant Professor 2005-2010 and Associate Professor 2010-2013 at Case Western Reserve University. Between her academic positions, she was a Senior Technologist at Western Digital/SanDisk Corporation. Dr. Zhang's research spans the areas of VLSI architecture design, digital storage and communications, security, and signal processing.

Dr. Zhang received an NSF CAREER Award in January 2009. She is also the recipient of the Best Paper Award at 2004 ACM Great Lakes Symposium on VLSI and 2016 International SanDisk Technology Conference. She authored the book ``VLSI Architectures for Modern Error-Correcting Codes'' (CRC Press, 2015), and co-edited ``Wireless Security and Cryptography: Specifications and Implementations" (CRC Press, 2007). She also published more than 100 papers. She was elected to serve on the Board of Governers of the IEEE Circuits and Systems Society for the 2019-2021 term and is a member of the CASCOM and VSA technical committees. She was also a Co-Chair of the Data Storage Technical Committee (2017-2020). She served on the technical program and organization committees of many conferences, including ISCAS, SiPS, ICC, GLOBECOM, GlobalSIP, and GLSVLSI. She has been an associate editor for the IEEE Transactions on Circuits and Systems-I 2010-2019 and IEEE Open Journal of Circuits and Systems since 2019.
\end{IEEEbiography}


\begin{thebibliography}{2}

\bibitem{Resnet}
K. He, X. Zhang, S. Ren, and J. Sun, ``Deep Residual Learning for Image Recognition,'' {\it Proceedings of IEEE Conference on Computer Vision and Pattern Recognition,}, pp. 770-778, 2016.

\bibitem{network express}
M. Raghu, B. Poole, J. Kleinberg, S. Ganguli, and J. S. Dickstein, ``On the expressive power of deep neural networks'', {\it Proceedings of the 34th International Conference on Machine Learning}, vol. 70, pp. 2847-2854, Aug. 2017.

\bibitem{Eyeriss}
Y. Chen, {\it et. al.}, ``Eyeriss: an energy efficient reconfigurable accelerator for deep convolutional neural networks'', {\it IEEE Journal of Solid-State Circuits}, vol. 52, no.1, pp. 127-138, 2017.

\bibitem{Sparten}
A. Gondimalla, {\it et. al.}, “Sparten: A sparse tensor accelerator for convolutional neural networks,” {\it Proceedings of 2019 IEEE/ACM Annual International Symposium on Microarchitecture}, pp. 151–165, 2017.

\bibitem{PermDNN}
C. Deng, S. Liao, and B. Yuan, ``PermDNN: energy-efficient convolutional neural network hardware architecture with permuted diagonal structure,'' {\it IEEE Transactions on Computers}, vol. 70, no. 2, pp. 163–173, 2021.

\bibitem{GOSPA}
C. Deng, {\it et. al.}, ``Gospa: an energy-efficient high-performance globally optimized sparse convolutional neural network accelerator'' {\it Proceedings of 2021 ACM/IEEE Annual International Symposium on Computer Architecture}, pp. 1110-1123, 2021.

\bibitem{ESCALATE}
S. Li, {\it et. al.}, ``Escalate: boosting the efficiency of sparse DNN accelerator with kernel decomposition'' {\it Proceedings of 2021 IEEE/ACM International Annual Symposium on Microarchitecture}, pp. 992-1004, 2021.

\bibitem{RoyEPIC}
J. A. Roy, F. Koushanfar, and I. L. Markov, ``EPIC: ending piracy of integrated circuits,'' {\it Proceedings of Conference on Design, Automation and Test in Europe}, pp. 1069-1074, 2008.

\bibitem{SAT}
P. Subramanyan, S. Ray, and S. Malik, “Evaluating the security of logic encryption algorithms,” {\it Proceedings of 2015 IEEE International Symposium on Hardware Oriented Security and Trust}, pp. 137–143, 2015.

\bibitem{RajendranSecurity}
J. Rajendran, Y. Pino, O. Sinanoglu, and R. Karri ``Security analysis of logic obfuscation,'' {\it Proceedings of ACM/IEEE Design Automation Conference}, pp. 83-89, 2012.

\bibitem{AND logic}
S. Dupuis, P. -S. Ba, G. Di Natale, M. Flottes, and B. Rouzeyre, ``A novel hardware logic encryption technique for thwarting illegal overproduction and hardware trojans,'' {\it IEEE International On-Line Testing Symposium} pp. 49-54, 2014.

\bibitem{Fault analysis}
J. Rajendran, {\it et. al.}, ``Fault analysis-based logic encryption,'' {\it IEEE Transactions on Computers,} vol. 64, no. 2, pp. 410-424, Feb. 2015.

\bibitem{Lee-logic}
Y. Lee and N. A. Touba, ``Improving logic obfuscation via logic cone analysis,'' {\it Proceedings of Latin-American Test Symposium,} pp. 1-6, 2015.

\bibitem{LUT2}
B. Liu and B. Wang, ``Embedded reconfigurable logic for ASIC design obfuscation against supply chain attacks,'' {\it Proceedings of Conference on Design, Automation and Test in Europe}, pp. 1-6, 2014.

\bibitem{Anti-SAT}
Y. Xie and A. Srivastava, “Anti-SAT: mitigating SAT attack on logic locking,” {\it IEEE Transactions on Computer-Aided Design Integration Circuits System} , vol. 38, no. 2, pp. 199–207, Feb. 2019.

\bibitem{SARLock}
M. Yasin, B. Mazumdar, J. J. V. Rajendran, and O. Sinanoglu, “SARLock: SAT attack resistant logic locking,” {\it Proceedings of 2016 IEEE International Symposium on Hardware Oriented Security and Trust}, pp. 236–241, 2016.

\bibitem{AppSAT}
K. Shamsi, {\it et. al.}, ``On the approximation resiliency of logic locking and IC camouflaging schemes,” {\it IEEE Transactions on Information Forensics Security,} vol. 14, no. 2, pp. 347–359, Feb. 2019.

\bibitem{G-Anti-SAT}
J. Zhou and X. Zhang, ``Generalized SAT-attack-resistant logic locking,'' {\it IEEE Transactions on Information Forensics and Security}, vol. 16, pp. 2581-2592, 2021.

\bibitem{ISCAS}
J. Zhou and X. Zhang, ``A new logic-locking scheme resilient to gate removal attack,'' {\it Proceedings of 2020 IEEE International Symposium on Circuits and Systems}, pp. 1-5, 2020

\bibitem{SFLL}
M. Yasin, {\it et. al.}, “Provably-secure logic locking: From theory to practice,” {\it Proceedings of the 2017 ACM SIGSAC Conference on Computer and Communications Security}, pp. 1601–1618, 2017.

\bibitem{removal}
M. Yasin, {\it et. al.}, “Removal attacks on logic locking and camouflaging techniques,” {\it IEEE Transactions on Emerging Topics in Computing}, vol. 8, no. 2, pp. 517-532, 1 April-June 2020.

\bibitem{ML}
Y. Liu, M. Zuzak, Y. Xie, A. Chakraborty, and A. Srivastava, ``Robust and attack resilient logic locking with a high application-level'' {\it ACM Journal on Emerging Technologies in Computing Systems,}  vol. 17, no. 3, pp. 1-22, July 2021.

\bibitem{FALL}
D. Sironee and P. Subramanyan, ``Functional analysis attacks on logic locking,'' {\it IEEE Transactions on Information Forensics and Security,} vol. 15, pp. 2514-2527, Jan. 2020.

\bibitem{HD-unlock}
F. Yang, M. Tang, and O. Sinanoglu, ``Stripped functionality logic locking With hamming distance-based restore unit (SFLL-hd) – unlocked,'' {\it IEEE Transactions on Information Forensics and Security,} vol. 14, no. 10, pp. 2778-2786, Mar. 2019.

\bibitem{ML-doc}
Y. Liu, {\it et. al.}, ``ML-Doctor: Holistic risk assessment of inference attacks against machine learning models,'' {\it Proceedings of 31st USENIX Security Symposium,} Aug., 2022. 

\bibitem{protection1}
A. Chakraborty, A. Mondai and A. Srivastava, "Hardware-assisted intellectual property protection of deep learning models," {\it Proceedings of 57th ACM/IEEE Design Automation Conference}, pp. 1-6, 2020.

\bibitem{protection2}
B. F. Goldstein, V. C. Patil, V. C. Ferreira, A. S. Nery, F. M. G. França and S. Kundu, "Preventing DNN model IP theft via hardware obfuscation," {\it IEEE Journal on Emerging and Selected Topics in Circuits and Systems}, vol. 11, no. 2, pp. 267-277, June 2021.

\bibitem{Alexnet}
A. Krizhevsky, I. Sutskever, and G. E. Hinton, ``ImageNet classification with deep convolutional neural networks'', {\it Proceedings of the 25th International Conference on Neural Information Processing Systems}, vol. 1, pp. 1097-1105, 2012.

\bibitem{VGG}
K. Simonyan and A. Zisserman, ``Very deep convolutional networks for large-scale image recognition,'' {\it Proceedings of the 3rd International Conference on Learning Representations,} 2015.

\bibitem{dataset}
J. Deng {\it et. al.}, ``Imagenet: A large-scale hierarchical image database'', {\it Proceedings of 2009 IEEE Conference on Computer Vision and Pattern Recognition}, pp.248-255, 2009.

\bibitem{CIFAR-10}
A. Krizhevsky, ``Learning multiple layers of features from tiny images,'' {2009}

\bibitem{DNN reliability}
B. F. Goldstein, {\it et. al.}, ``Reliability evaluation of compressed deep learning models,'' {\it Proceedings of 2020 IEEE Latin American Symposium on Circuits and Systems}, pp. 1-5, 2020.

\bibitem{Fashion}
Fashion-MNIST. https://github.com/zalandoresearch/fashion-mnist.

\end{thebibliography}
\end{document}